\documentclass[aps,prl,twocolumn,showpacs,showkeys,superscriptaddress,groupedaddress]{revtex4}  
\usepackage{graphicx, epstopdf}  
\DeclareGraphicsExtensions{.ps}
\usepackage{dcolumn}   
\usepackage{bm}        
\usepackage{amssymb}   
\usepackage{graphicx}

\begin{document}



\title{Processing In-memory realization using Quantum Dot Cellular Automata }
\centerline{}
%

%
\author{P.P.~Chougule} \affiliation{Computational Electronics and Nanoscience Research Laboratory,
School of Nanoscience and Biotechnology, Shivaji University, Kolhapur- 416004, India}
\author{B.~Sen} \affiliation{Department of Computer Science and Engineering,
National institute of Technology, Durgapur, W.B- 713209, India}
\author{T.D.~Dongale} \affiliation{Computational Electronics and Nanoscience Research Laboratory,
School of Nanoscience and Biotechnology, Shivaji University, Kolhapur- 416004, India}

%
%


\begin{abstract}
The present manuscript deals with the realization of Processing In-memory (PIM) computing architecture using Quantum Dot Cellular Automata (QCA) and Akers array. The PIM computing architecture becomes popular due to its effective framework for storage and computation of data in a single unit. Here, we illustrate two input NAND and NOR gate with the help of QCA based Akers Array as case study. The QCA flip flop is used as a primitive cell to design PIM architecture. The results suggested that, both the gate have minimum power dissipation. The polarization results of proposed architecture suggested that the signals are in good control. The foot print of the primitive cell equals to 0.04  $\mu m^2$
, which is smaller than conventional CMOS primitive cell. The combination of QCA and Akers array provides many additional benefits over the conventional architecture like reduction in the power consumption and feature size, furthermore, it also improves the computational speed.
\end{abstract}



\maketitle{}

\section{I. Introduction}

         It is a well established fact that the traditional CMOS technology has facing the wall in terms of further scaling down besides other issues such as power dissipation, speed and footprint. This has posed a bottleneck for the conventional von-Neumann architecture which comprises of the well known two units with well defined tasks viz. processing and storing of data. The basic notion of von-Neumann and PIM computing is shown in fig. 1 for the ready reference and also for setting the background of our development. As against the requirement of two separate units for processing and storing of the data in the conventional von-Neumann, PIM computing architecture does the same only with one unit. This will be benefited in terms of speed, feature size and power consumption. There are various way to implement the In-memory computing architecture but combination of Quantum Dot Cellular Automata (QCA) and Akers array provides additional benefits over other counterparts.  
                      
         The QCA is a promising and reliable technique for nanoelectronics devices and architecture [1-2]. The Akers array is known for its In-memory computing capabilities hence the combination of Akers array and QCA can be a great hope for the future computing architecture. Recently, Awais et al have reported a low density parity check (LDPC) decoding algorithm using QCA. There results suggested that QCA based LDPC is more area efficient than complementary metal oxide semiconductor (CMOS) technology [3]. Retallick et al have embedded QCA circuits onto a quantum annealing processor. They have uses dense placement and heuristic algorithm to characterize the QCA circuits [4]. Moustafa et al have reported the QCA for synthesis of classical and reversible circuits. They have realized XOR, XNOR, CNOT and Toffoli gates using QCA [5]. Angizi et al have reported QCA based RAM cell. The proposed design was based on the majority gate concept. They have also employ SET and REST functionality to QCA based RAM cell [6]. Though there are other PIM techniques available like memristive Akers array, the lower feature size along with reduced power consumption of QCA cells serves as the deciding factors for the PIM architecture [7]. 
         
         In our previous study, we have constructed PIM architecture by employing QCA multiplexer [8]. In the QCA multiplexer based PIM architecture, there is less control on signal. So, we intend to combine of Akers array and QCA flip-flop so as to synergize their capabilities for improving the computing metrics. We present through this paper, an approach of PIM computing using QCA flip-flop based Akers Array. Rest of the paper is organized as follows, after general introduction; second section covers the overview of Akers array and QCA. The proposed QCA flip-flop based Akers array is reported in the third section. This is followed by the case study of two input NAND and NOR gate. At the end results and discussion are presented.

\begin{figure}[h]
	\includegraphics[width=8cm]{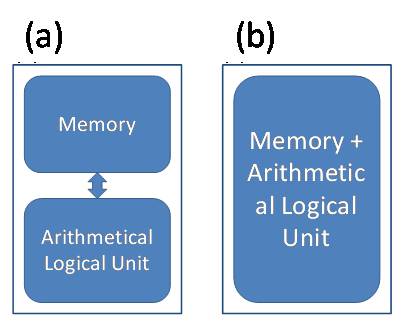}
	\caption{ Computer architectures. (a) von-Neumann architecture: separate memory and processing unit. (b) In-memory computing architecture: both memory and processing unit placed in single unit. }
\end{figure}

\section{II. Background}
The present section deals with the background of Akers logic array and QCA. This is followed analytical combination of above with the QCA followed by the details of cell structure and functional operation.

\subsection{Akers logic array }
     In 1972, S. B. Akers proposed a new rectangular logic array, which has in-memory computing capabilities. Fig. 2 represents an Akers logic array, it is a two dimensional array of identical logical cells connected in rectangular grid. The basic function of the array is governed by following function: [9]
     
     \begin{equation}
     F(X, Y, Z) = X(\bar{Z}) +YZ	
     \end{equation}									
     
     In original publication of Akers logic array, four alternative logical operations have been presented that generates the correct behavior of array [9]. The unit cell of Akers array consist of three inputs 'X', 'Y', and 'Z' and produce two outputs 'F' of same value as shown in the fig. 2(a). Fig. 2(b) represents 3x3 input's two dimensional Akers array. Here 'X' and 'Y' are considered as binary input terminals where as 'Z' considered as control input terminal which can be useful for the storing of current state of the QCA [7]. The output of every logic cell in array is transferred to the next two neighboring logic cells; one is given to lower and other being at the right neighbor cell. The data is then transferred to vertical and horizontal neighbors in array, as shown in fig. 2(b). The input ‘X’ of the uppermost row is set to ‘zero’ for all columns and the input ‘Y’ of the column is set to ‘one’ for all rows. The execution of Boolean function is performed by arranging the array cells according the specific manner. We get output function at the lower right cell of array. The same array can be used for different Boolean functions, each specifying different arrangement of inputs. The prototype examples of different Boolean functions such as four inputs XOR gate is shown in fig. 3. The control terminal 'Z' stored the data where as entire array perform the Boolean functions hence the Akers array considered as the PIM computational structure.  

\begin{figure}[h]
	\includegraphics[width=8cm]{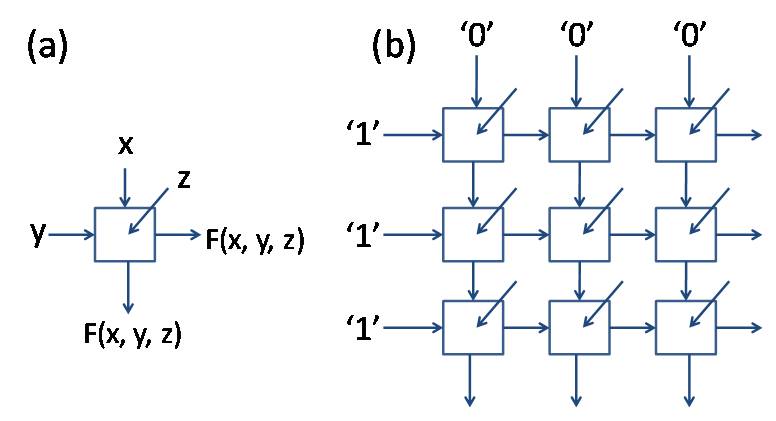}
	\caption{ Akers array (a) A logic cell with three inputs X, Y, and Z and two identical outputs (b) Akers 3x3 array structure [9]. }
\end{figure}

\begin{figure}[h]
	\includegraphics[width=8cm]{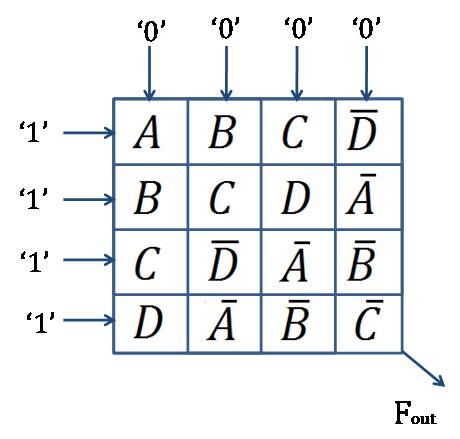}
	\caption{ Examples of four input XOR gate [7]. }
\end{figure}

\subsection{Quantum dot Cellular Automata (QCA)  }
QCA is a cellular automata with unique cells. Through the present communication in fact we are showcasing the same, as an alternative for conventional CMOS technology. The cells of QCA consist of four quantum dots and two diagonally sited electrons [10], as shown in fig. 4. 

\begin{figure}[h]
	\includegraphics[width=8cm]{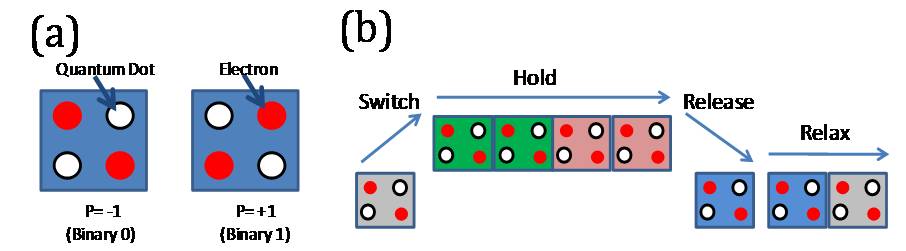}
	\caption{ Examples of four input XOR gate [7](a) QCA cell structure. It consists of four quantum dots and two electrons. The output depends upon the diagonal position of electrons. (b) Clocks in QCA and its four phases. }
\end{figure}

        As mentioned above, the QCA cells comprises of two electrons in the diagonal position owing to the columbic repulsive force. So, the possibility of electrons arrangement in a cell resembles to only two. First diagonal arrangement is considered as logical ‘0’ which has polarity -1 while the second diagonal being logical ‘1’ with  polarity +1, as shown in fig. 4(a). In QCA structure, the output of first QCA cell acts like input for next cell while the second cell produce the same output similar to first cell due to the stability of electrons. The same is carried forward to the next cell. Due to this reason, QCA cell can possible be used to make wires and forms the basis of many other Boolean functions by arranging them in explicit manner. 
        
        Columbic interaction between adjacent cells makes them align in same polarization or same state. This is the main principle behind the QCA wire. For the more complicated circuit there was a need of control on the direction of information. Finally, clocked QCA cells came in the picture. Now clocking became the undividable part of QCA. Generally, every clock has four phases viz. switch, hold, release, and relax. Depending on these four phases, there are four clocks (clock 0, clock 1, clock 2, and clock 3) required to control the flow of information through QCA cells. A brief information of every state of clock is as follows:
        
        \textbf{1. Switch state-} At this state electrons are allowed to tunnel from one quantum dot to adjacent quantum dot in single cell. Here electrons columbic repulsion force is greater than the potential energy of quantum dot. So, the higher repulsion energy allows the electron to tunnel from one quantum dot to other. This is the reason that QCA cell can change its polarization as per its neighbor cell.
        
        \textbf{2. Hold state-} Here electrons are not allowed to tunnel between quantum dots. The potential of quantum dot is much higher than the repulsion force between electrons. Due to this reason, electrons get localized in quantum dot. This strict localization of electrons proves, QCA possess a non-volatile memory. 
        
        \textbf{3. Release state-} In release state, electrons are free to tunnel. This state looks similar to switch state but it is in the other part of cycle. In the release state, electrons tunnel from some fixed polarity, whereas in switch state electron tunnel from the null point.
        
        \textbf{4. Relax state-} At this stage, there is no fixed polarity present nor the electron tunneling happens. They are neither localized state nor tunneling state. They don’t show any logical output, hence this state is called as ‘relax state’.
         
        After the relax state, QCA cell again get the switching state and the clocking cycle repeats once again. Here the important rule is that, when clock 0 is in the switching state, clock 1 will be in hold state, clock 2 release state and clock 3 is in relax state. These clocks helps to control the flow of information in QCA system.

\section{III. Proposed QCA Akers logic array}
         The implementation of Akers logic array in conventional CMOS technology is impractical due to the inherent outsized silicon estate [7]. Hence we are proposing QCA based system owing to its atomic dimension. Consequently QCA based Akers logic array has proved to be many times denser than conventional CMOS, which significantly reduces silicon real estate. The proposed QCA-Akers logic array cells are designed using QCA flip-flop. The structure and simulation of QCA flip-flop is shown in the fig. 5 (a and b) respectively. 

\begin{figure}[h]
	\includegraphics[width=8cm]{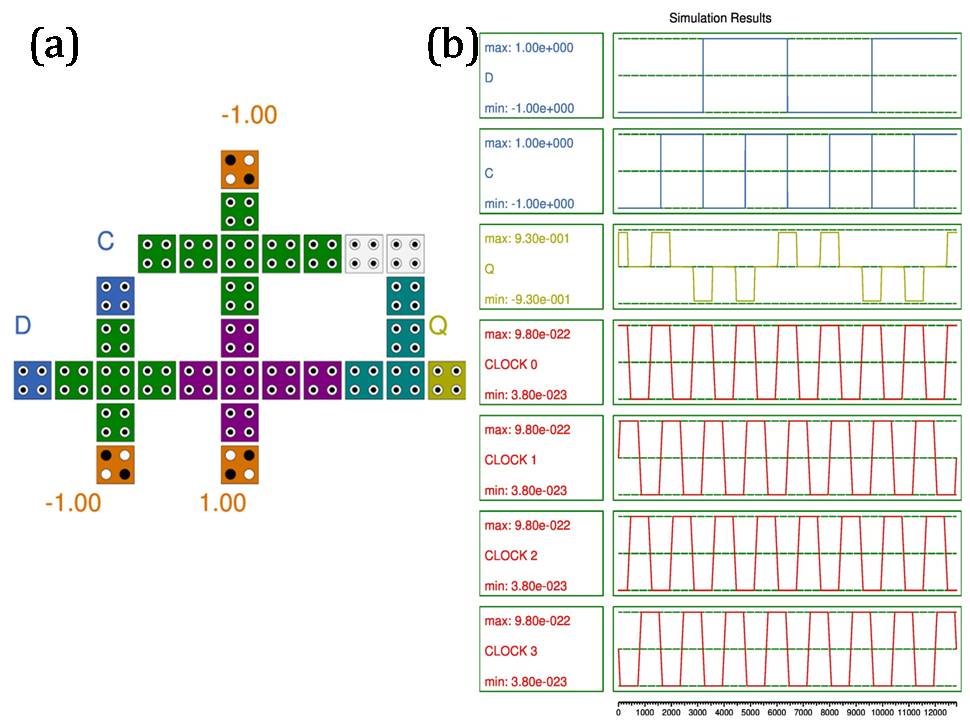}
	\caption{ (a) QCA Flip-Flop [11]. (b) simulation results of QCA Flip-Flop. }
\end{figure}

\subsection{Structure of primitive logic cell}
The primitive cell consists of two complimentary QCA flip-flop, as illustrated in the fig. 6 (a). The control inputs of cell ‘X’ and ‘Y’ are given as fixed input i.e. zero and one respectively. The control input ‘Z’ is used for storing the logical state of QCA cell i.e. Qz, which is represented by the clock of that circuit. The stored logical state of Qz and Qz ̅  are written during write operation anterior to execution. 

\begin{figure}[h]
	\includegraphics[width=8cm]{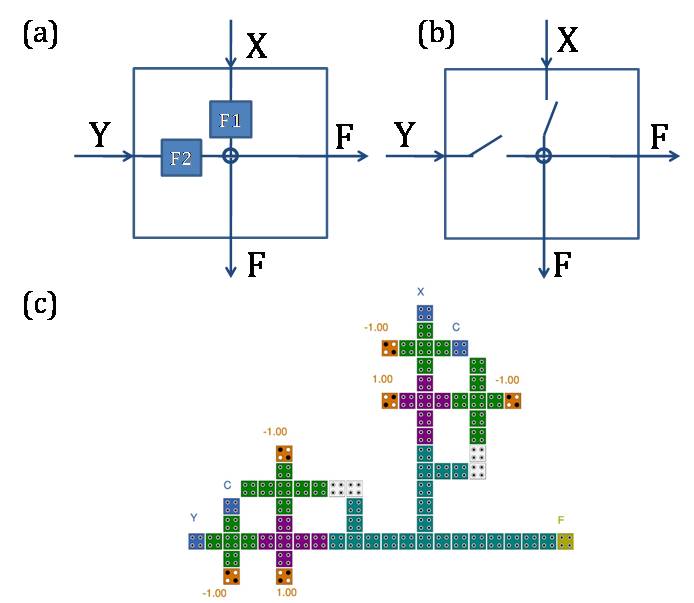}
	\caption{ Primitive logic cell. (a) Schematic diagram of proposed primitive logic cell. (b) A behavioral model of the basic logic cell, where QCA Flip-Flops are used as switches. (c) Proposed primitive logic cell by using QCA. }
\end{figure}

         Ideally QCA Flip-Flops can be function as switches. When clock 'C' becomes 1, then it act as closed circuit , otherwise open circuit as shown in fig. 6 (b). In actual Akers model, if one switch is open then other is closed. By using these control inputs, the desired output from the Akers logic cell can be derived. The arrangement of QCA cells in the Akers cell satisfies the functionality as mentioned in the section 2.1.
         
\subsection{Logic Array Operation}

The Akers Logic array is a memory array with additional computational capabilities. The Akers array logic is known as In-memory logic. This array can compute all Boolean functions in additional to storing data. The computational operation in Akers Logic array is divided into two stages. The initial stage is ‘write’ operation to the QCA Akers cells. In this particular stage, the initial stored logical state of QCA Akers cells Qz and Qz ̅ is written. This stage can be expressed as the initial state of regular write operation of memory or alternatively as the base of computing Boolean function for next stage [7]. 

\textbf{(A) Stage 1-Initialization of primitive logic cell (write operation)}
Initialization of primitive logic cell of logic states Qz and Qz ̅ is simultaneously achieved by connecting both QCA Flip-Flops in complimentary manner, as illustrated in fig. 6(a). They can be made complimentary by controlling the clock signal of each Flip-Flop. In this complimentary structure, one Flip-Flop has active clock signal and other has inactive clock signal. 

To write logical 1 to Qz, the clock signal of Qz and Qz ̅ should be active and inactive respectively. To write 0 to Qz, the arrangement should be exactly apposite of first. Therefore, by only controlling the clock signal of Flip-Flop, one can control the output of each QCA Akers cell. Fig. 7 (a and b) represents an illustrative example of three input XOR gate using QCA-Akers array and it behavioral model respectively.

\begin{figure}[h]
	\includegraphics[width=8cm]{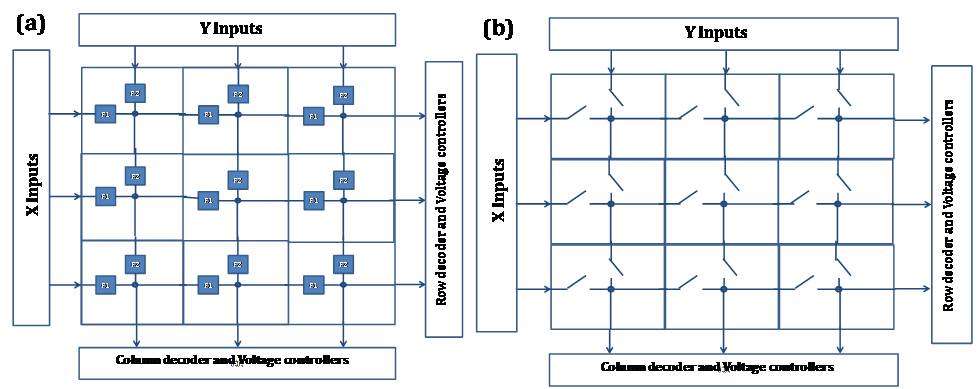}
	\caption{(a) Three input XOR gate using QCA-Akers array. (b) The behavioral model of three input XOR gate, where QCA cells are used as switch [7]. }
\end{figure}

\textbf{(B) Stage 2-Execution of primitive logic cell (read operation)}
To maintain the correct operation of QCA Akers logic array, the two clocks should not be active in single primitive cell. By maintaining clock signals, we can get output of the basic primitive Akers logic cell. The output transfers to neighboring cells. The rightmost lowest cell’s output is the final output. So, we get final result at rightmost lowest cell.

\section{IV.Implementation Details of In-memory Architecture}
To evaluate QCA Akers logic array, two Boolean functions were investigated. In this section, working and logical principle is described considering two test examples viz. two input NAND gate and NOR gate.

\subsection{QCA Akers NAND gate}
The proposed QCA-Akers based NAND gate is shown in the fig. 8. It consists of two Akers cells arranged horizontally and an inverter afterward. One of the advantages of this structure is that the cell count increases in arithmetic progression whereas for XOR gate it increases in geometric progression. Hence for the higher number of inputs the overall cell count is very small as compared to XOR gate.   

\begin{figure}[h]
	\includegraphics[width=8cm]{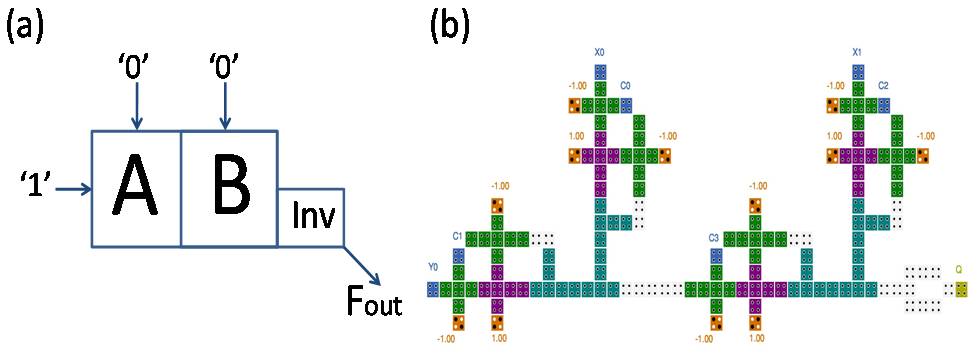}
	\caption{Two input NAND gate. (a) Structure of Akers two input NAND gate. (b) QCA Akers two input NAND gate. }
\end{figure}

The proposed NAND gate consists of two Akers cells which are connected in parallel manner followed by an inverter. Each cell has capability to store as well perform computation on data. Here we provided ‘X’ and ‘Y’ as a fixed inputs and we can control the value of ‘Z’ i.e. logical state of the QCA cell. There are only four inputs combination available for NAND gate viz. '00', '01', '10', '11'. To satisfy the functionality of QCA Akers NAND gate, one has to satisfy every respective output of these four inputs.

\begin{figure}[h]
	\includegraphics[width=8cm]{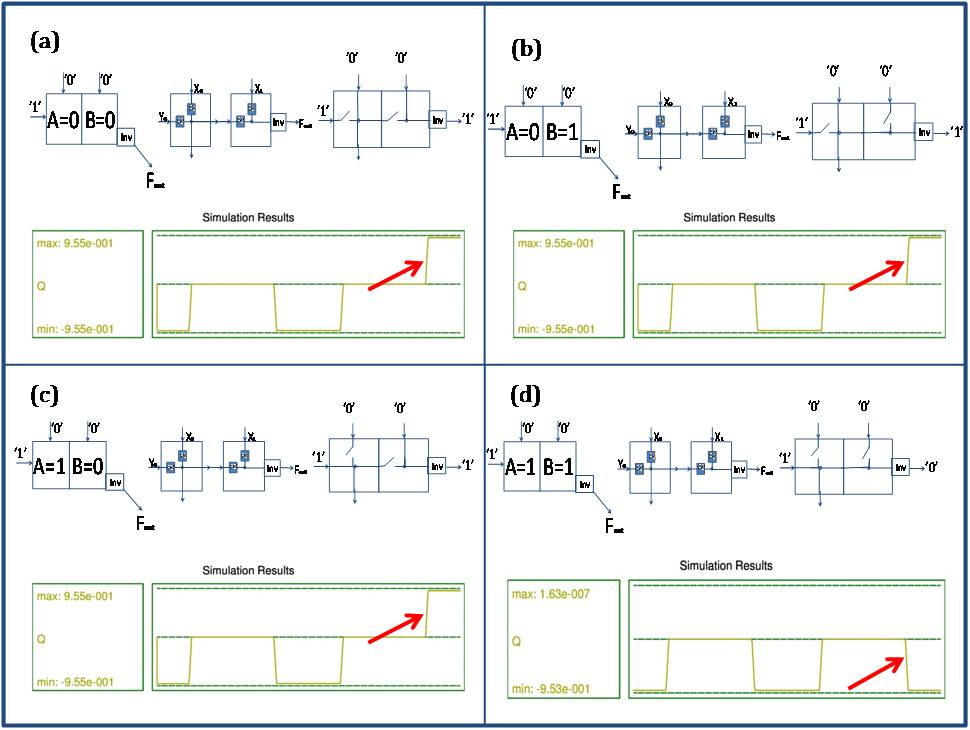}
	\caption{Simulation results of QCA Akers two input NAND gate. }
\end{figure}

Fig. 9 (a to d) represents the simulation results of QCA Akers NAND gate. Fig. 9(a) represents the case when, both the inputs are zero. In this case only Y0 line’s switches are open, whereas X0 and X1 lines switches become close. Due to this the output signal at lowest right most corner becomes ‘1’. This satisfies the first case of NAND gate. For the second case of input ‘01’, the X0 line acts as close switch whereas X1 and Y0 will act as a open switches. This combination of states create a logic ‘1’ output. For the third case of input ‘10’, X1 line’s switch acts as a close switch. So, we get ‘0’ signal at the output stage but an inverter makes it logic ‘1’. For the last state of input ‘11’, Y0 lines switches work as closed switches hence its output transferred to Fout line which is nothing but logic ‘0’. In this way, the QCA based Akers logic array can perform NAND logic functions and the control signal i.e. clock signal of QCA will be responsible for the storing of intermediate data.

\subsection{QCA Akers NOR gate}
The proposed QCA-Akers NOR gate consist of two Akers cells arranged vertically and an inverter is placed after lower cell. Here Flip-Flops are used as switches as well as every Flip-Flop stores one bit information. This property of Flip-Flop is useful for the In-memory computing application. The proposed QCA-Akers NOR gate is shown in the fig. 10. In the present case, ‘X’ and ‘Y’ are considered as fixed inputs whereas ‘Z’ worked as control input. To describe working of QCA Akers NOR gate, one has to satisfy every possible output of four inputs. 

\begin{figure}[h]
	\includegraphics[width=8cm]{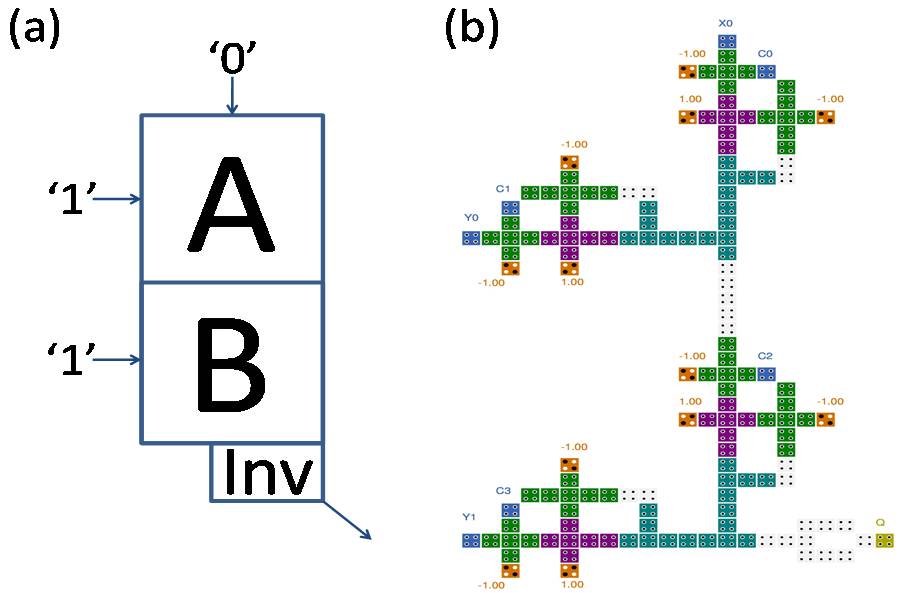}
	\caption{Two input NOR gate. (a) Structure of Akers two input NOR gate. (b) QCA Akers two input NOR gate. }
\end{figure}

Simulation results of QCA Akers two input NOR gate are shown in the fig. 11 (a to d). Fig. 11 (a) represents the case when both the inputs are at zero state and X0 line’s switches are worked as closed switches. In this case, X0 signal i.e. ‘0’ moves forward and after inverter block it becomes ‘1’. This satisfies the first case of NOR gate. For the second case of input ‘01’, the Y0 line acts as open switch where as Y1 will act as a close switch followed by an inverter. This will make output equals to logic ‘0’. For the third case of input ‘10’, X0 and Y1 line’s switch acts as an open switch and Y0 acts as closed switch. This will make output equals to logic ‘0’. For the last state of input ‘11’, Y0 and Y1 line’s switches work as closed switches. Due to this, the output transferred to Fout line which is nothing but logic ‘0’. In this way, the QCA based Akers logic array can perform NOR logic functions and the control signal i.e. clock signal of QCA will be responsible for the storing of intermediate data.

\begin{figure}[h]
	\includegraphics[width=8cm]{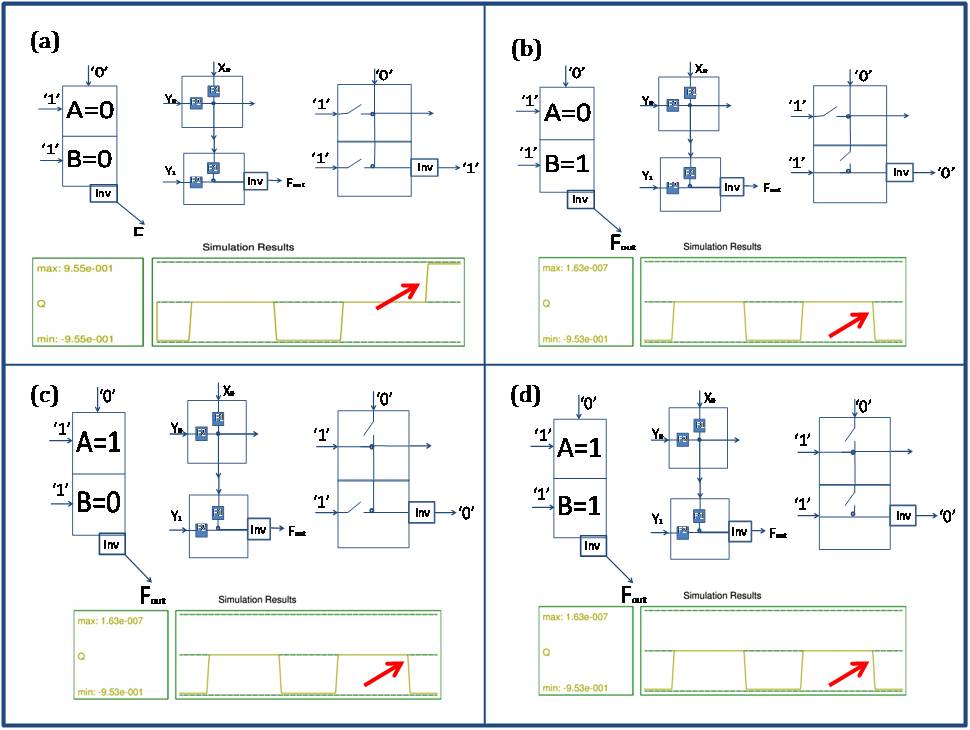}
	\caption{Simulation results of QCA Akers two input NOR gate. }
\end{figure}

\section{V. Result and Discussion}
The architecture presented in this paper comprises two complementary flip-flop switches in every primitive cell. This structure is very useful for PIM computing applications. The advantage of this circuit is that, every primitive cell as whole can store one bit of memory. Every primitive cell contains two flip-flops, hence ultimately a single primitive cell can store two bit of memory as well as can perform logic operation on the data. The NAND and NOR gate are considered as universal gates hence we can design different kind of logic architecture using these universal gates. The proposed architecture has very less feature size and power dissipation. The power dissipation analysis of NAND and NOR gate is summarized in the table 1. Furthermore the power dissipation map of the proposed gates is shown in the fig. 12.

\begin{figure}[h]
	\includegraphics[width=8cm]{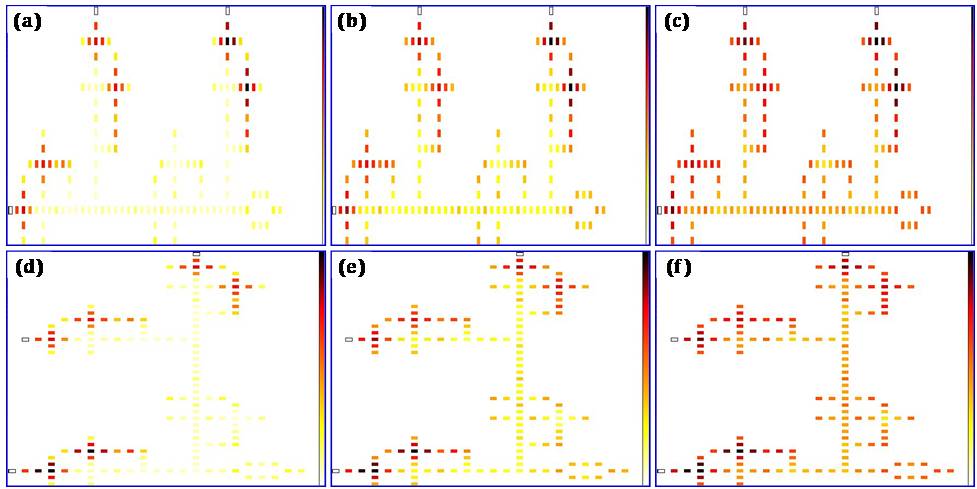}
	\caption{Power dissipation map of QCA Akers NAND and NOR gate. Power dissipation map at (a) Ek=0.5 meV; (b) Ek=1.0 meV; (c); Ek=1.5 meV; (d) Ek=0.5 meV; (e) Ek=1.0 meV; (f) Ek=1.0 meV. }
\end{figure}

\begin{figure}[h]
	\includegraphics[width=8cm]{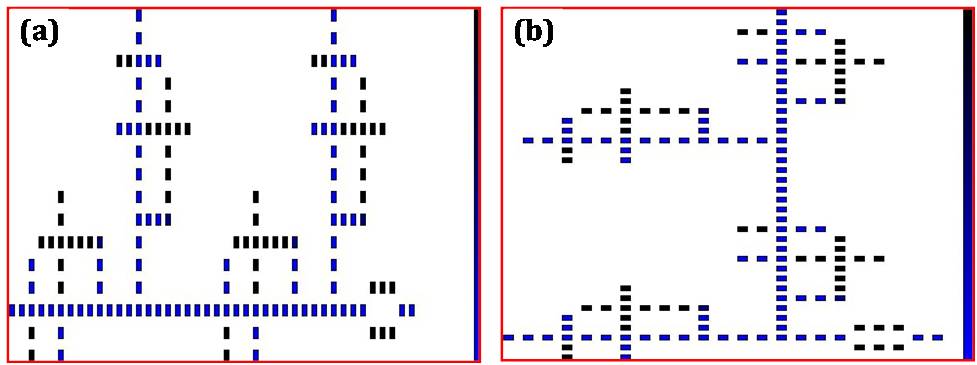}
	\caption{(a and b): Polarization information of the NAND and NOR gate respectively. }
\end{figure}

The results suggested the power dissipation of both the circuit becomes maximum at tunneling energy Ek equal to 1.5 meV. The polarization results of proposed architecture is given in the fig. 13 (a and b). The polarization effect will make the signal strong in the circuit. The results suggested that the proposed circuit have good control over the signals. The QCA flip-flop comprises 30 QCA cells and its footprint equals to 0.04  $\mu m^2$. The presented NAND and NOR gate consist of 147 QCA cells and its total footprint equals to 0.31  $\mu m^2$ and 0.34  $\mu m^2$ respectively, which is way smaller than conventional CMOS NAND and NOR gate. The overall results suggested that, the proposed architecture is more efficient than the conventional CMOS gates. The combination of QCA and Akers array provides many additional benefits over the conventional computer architecture like reduction in the power dissipation and feature size, furthermore, it also improves the computational speed.

\begin{table}
	\centering
	\caption{Power dissipation analysis of NAND and NOR gate}
	\begin{tabular}{|l|l|l|l|l|l|l|}
		\hline
		Parameter                              & Ek =0.5 (meV) & ~           & Ek =1.0(meV) & ~            & Ek =1.5 (meV) & ~           \\ \hline
		~                                      & NAND          & NOR         & NAND         & NOR          & NAND          & NOR         \\ 
		Max Kink Energy                        & 0.00148 meV   & 0.00148 meV & 0.00148 meV  & 0.00148 meV  & 0.00148 meV   & 0.00148 meV \\ 
		Max Energy dissipation of circuit      & 0.20704 meV   & 0.21153 meV & 0.27689 meV  & 0.28150 meV  & 0.36453 meV   & 0.36934 meV \\ 
		Max Energy dissipation vector          & 20            & 20          & 20           & 20           & 20            & 20          \\ 
		Average Energy dissipation of circuit  & 0.11131 meV   & 0.11366 meV & 0.19465 meV  & 0.19734 meV  & 0.29388 meV   & 0.29706 meV \\ 
		Max Energy dissipation among all cells & 0.00785 meV   & 0.00785 meV & 0.00735  meV & 0.00735  meV & 0.00707 meV   & 0.00707 meV \\ 
		Max Energy dissipation vector          & 31            & 31          & 31           & 31           & 31            & 31          \\ 
		Min Energy dissipation of circuit      & 0.04493 meV   & 0.04519 meV & 0.13668 meV  & 0.13754 meV  & 0.24430 meV   & 0.24590 meV \\ 
		Min Energy dissipation vector          & 00            & 00          & 00           & 00           & 00            & 00          \\ 
		Average Leakage Energy dissipation     & 0.04629 meV   & 0.04653 meV & 0.13869 meV  & 0.13953 meV  & 0.24638 meV   & 0.24797 meV \\ 
		Average Switching Energy Dissipation   & 0.06502 meV   & 0.06713 meV & 0.05596 meV  & 0.05781 meV  & 0.04750 meV   & 0.04909 meV \\
		\hline
	\end{tabular}
\end{table}

\end{document}